\documentclass[12pt,reqno,showkeys]{amsart}

\usepackage{amsthm,amsmath,amsfonts,amssymb}
\usepackage{mathrsfs}
\usepackage{ascmac}
\usepackage{bm}
\usepackage{natbib}
\usepackage{here,comment}
\usepackage{ascmac}
\usepackage{fancybox}
\usepackage{graphicx}
\usepackage{xcolor}
\usepackage{hyperref}

\usepackage{multirow}
\usepackage{siunitx}
\usepackage{booktabs}

\usepackage{amsaddr}
\usepackage{fancyhdr}
\usepackage{natbib}
\usepackage{fullpage}
\usepackage{newtxtext,newtxmath}

\theoremstyle{definition}

\newcommand{\R}{\mathbb{R}}

\newcommand{\mD}{\mathcal{D}}

\newcommand{\mS}{\mathcal{S}}
\newcommand{\mT}{\mathcal{T}}

\renewcommand{\hat}{\widehat}

\newcommand{\argmin}{\operatornamewithlimits{argmin}}

\DeclareMathOperator{\vect}{vec}

\usepackage{amsaddr}

\title{Reconciling Functional Data Regression with Excess Bases}

\author[1]{Tomoya Wakayama}
\address{Graduate School of Economics, The University of Tokyo\\ 7-3-1 Hongo, Bunkyo-ku, Tokyo, Japan}
\author[2]{Hidetoshi Matsui}
\address{Faculty of Data Science, Shiga University\\
1-1-1 Banba, Hikone, Shiga, Japan}

\date{\today, \textit{Contact}: \textit{tom-w9@g.ecc.u-tokyo.ac.jp}}

\begin{document}

\maketitle

\begin{abstract}
As the development of measuring instruments and computers has accelerated the collection of massive amounts of data, functional data analysis (FDA) has experienced a surge of attention. The FDA methodology treats longitudinal data as a set of functions on which inference, including regression, is performed. Functionalizing data typically involves fitting the data with basis functions. In general, the number of basis functions smaller than the sample size is selected. This paper casts doubt on this convention. Recent statistical theory has revealed the so-called double-descent phenomenon in which excess parameters overcome overfitting and lead to precise interpolation. Applying this idea to choosing the number of bases to be used for functional data, we show that choosing an excess number of bases can lead to more accurate predictions. Specifically, we explored this phenomenon in a functional regression context and examined its validity through numerical experiments. In addition, we introduce two real-world datasets to demonstrate that the double-descent phenomenon goes beyond theoretical and numerical experiments, confirming its importance in practical applications.
\\

\end{abstract}

\section{Introduction}

Functional data analysis (FDA) has emerged as a powerful tool for analyzing longitudinal data across diverse fields, including biology, medicine, economics, and the social sciences~\citep{ramsay2005functional, horvath2012inference, kokoszka2017introduction, wang2016functional}. The fundamental concept of FDA is to represent the longitudinally measured data for each individual as a smooth function and then analyze the collection of functions using various statistical techniques~\citep{hsing2015theoretical}. This approach offers several advantages, such as reducing observational errors through smoothing and accommodating varying time points and numbers of observations for different subjects~\citep[e.g.,][]{Wakayama2024}.

In FDA, basis expansion is a widely used technique for transforming longitudinal data into functional data~\citep{fujii2006nonlinear,araki2009functional}. Basis expansion is known for its ability to smooth noisy data and reveal the underlying structure~\citep{green1994nonparametric, hastie2009elements}. In numerous FDA methodologies, such as functional regression and time series analysis, selecting the number of basis functions is a pivotal issue due to its substantial impact on prediction accuracy. The number of bases is selected from a range of values smaller than the number of observation points using information criteria \citep{akaike1973information,schwarz1978estimating,konishi1996generalised} or by employing cross-validation \citep{stone1974cross}. This practice aims to avoid overfitting, i.e., it seeks to mitigate the explosion of interpolated values between observation points. However, recent developments in statistical theory suggest that this approach may need to be reconsidered to achieve better prediction performance.

\begin{figure}[t]
  \begin{center}
  \includegraphics[width=\textwidth]{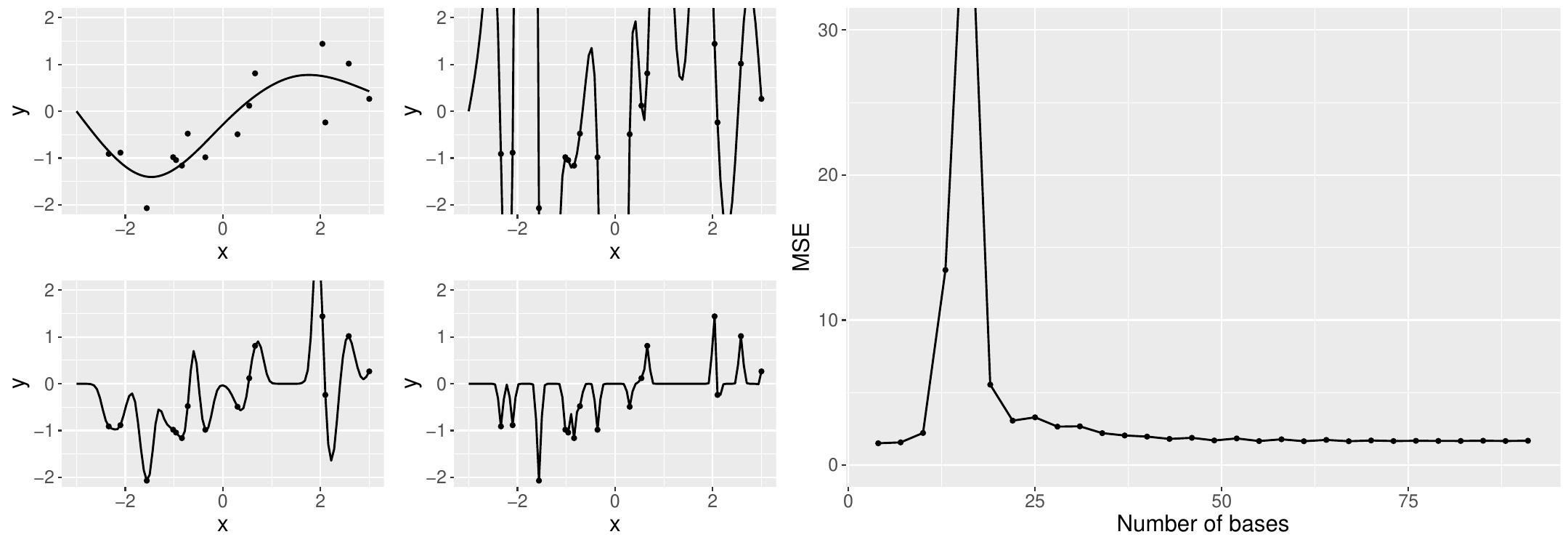}
  \end{center}
  \caption{Left: Curve fits when the number of
bases is $4$ (upper left), $20$ (upper right), $40$ (lower left), and $120$ (lower
right). Right: MSE for varying number of bases.}
  \label{fig:DDexample}
\end{figure}

Overfitting has long been a challenge in FDA; however, recent statistical theory has begun to reconcile this issue. Indeed, \cite{zhang2021understanding} empirically showed that deep neural network models with a large number of parameters that perfectly fit the training data can yield near-optimal accuracy for the test data. This phenomenon is referred to as the double-descent phenomenon~\citep{belkin2018understand,belkin2019reconciling}, where the interpolation error follows a conventional U-shaped curve up to a threshold, but decreases after reaching a peak at the threshold. In addition, \cite{hastie2022surprises,belkin2020two} theoretically revealed that the double-descent phenomenon can occur for linear regression models in several situations and showed the phenomenon empirically. For more detailed explanations, see \cite{james2013introduction, schaeffer2023double, misiakiewicz2023six} and references therein. Further, \cite{james2013introduction} demonstrated the double-descent phenomenon through a simple spline fitting. Figure~\ref{fig:DDexample} illustrates the phenomenon through fitting curves with measurement points. The figures on the left depict $15$ numerically generated data points and the spline curves fitted with the minimum norm interpolator~\citep{hastie2022surprises,bartlett2020benign} to estimate the parameters in the model for four different numbers of basis functions. A detailed description of the methodology is referred to in Section~\ref{sec:fun}. The right panel displays the mean squared errors in relation to the number of basis functions. When the number of bases equals the number of measurements, the spline curve appears overly undulating, which causes the mean squared error to explode. However, as the number of bases increases, the fitted curve becomes less undulating and the mean squared error decreases again. This suggests that using a large number of basis functions, especially a number larger than the sample size, may improve the accuracy of functional data analysis techniques.

In this paper, we advocate the use of a large number of basis functions, in combination with the minimum norm interpolator, to transform observed longitudinal data into functional data. Additionally, we apply the minimum norm interpolator to estimate functional regression models, which represent relationships between predictors and responses, either or both of which are given as functional data. We discuss four representative functional data regression scenarios where double descent is particularly relevant. We examine the effectiveness of the proposed approach within the four scenarios through simulation studies and applications to real-world datasets.

The remainder of the paper is organized as follows. Section~\ref{sec:fun} introduces functionalization with an excess number of basis functions. In Section~\ref{sec:method}, we discuss regression methods for functional data and their relation to the double-descent phenomenon. We validate our approach through numerical experiments in Section~\ref{sec:num}. Section~\ref{sec:app} demonstrates the importance of our advocations through applications to real datasets. Finally, we summarize our main points and suggest future research directions in Section~\ref{sec:con}.

\section{Functionalization} \label{sec:fun}
Functionalization is a crucial first step in functional data analysis. Without appropriate functionalization, extracting meaningful descriptive statistics or reaching accurate inferential conclusions becomes challenging in regression and classification. The process of functionalization involves transforming discrete, noise-corrupted observations into smooth functions that capture the underlying patterns and trends in the data \citep{ramsay2005functional}.

Suppose we have $N$ sets of time-course observations, where the $i$-th subject has $M_i$ observations $\{x_{i1}, x_{i2}, \ldots, x_{iM_i}\}$ at time points $\{t_{i1}, t_{i2}, \ldots, t_{iM_i}\}$ $(i=1, 2,\ldots, N)$, respectively, and $t_{ij}$ are elements of a domain $\mathcal T \subset \R$. We then consider transforming the time-course data into functions using the basis expansions \citep{ramsay2005functional, wang2016functional}. Let $\{\phi_k:\mT\to \R\}_{k=1}^K$ be a set of $K$ basis functions. We assume that each observation $x_{ij}$ can be expressed by the following regression form:
\begin{equation}\label{func-model}
    x_{ij} = \sum_{k=1}^Kw_{ik}\phi_k(t_{ij}) + \varepsilon_{ij} = \bm w_i^\top\bm\phi(t_{ij}) + \varepsilon_{ij} \quad (j=1, \ldots, M_i),
\end{equation}
where $\bm{w}_{i} = (w_{i1},w_{i2},\ldots,w_{iK})^{\top}$ is a vector of coefficients, $\bm{\phi}(t) =(\phi_1(t),\phi_2(t),\ldots,\phi_K(t))^{\top}$ is a vector of basis functions, and $\varepsilon_{i1}, \ldots, \varepsilon_{iM_i}$ are independent noise terms with mean $0$ and variance $\sigma_i^2$. Common choices for basis functions include the Fourier basis, spline basis, and wavelet basis~\citep{ramsay2005functional}.

We then calculate the optimal coefficient vector $\bm w_i$. Using the notation $\bm x_i = (x_{i1}, x_{i2}, \ldots, x_{iM_i})^\top$, $\Phi = (\bm\phi(t_{i1}), \bm\phi(t_{i2}),\ldots, \bm\phi(t_{iM_i}))^\top$, and $\bm\varepsilon_i = (\varepsilon_{i1}, \varepsilon_{i2}, \ldots, \varepsilon_{in_i})^\top$, the regression model~\eqref{func-model} can be expressed as $\bm x_i = \Phi\bm w_i + \bm\varepsilon_i$. We estimate $\bm w_i$ using the minimum norm interpolator~\citep{hastie2022surprises,bartlett2020benign}:
\begin{align*}
    \argmin_{\bm{w}_i\in \R^K }  \|\bm w_i\| \quad  {\rm s.t. } \quad \bm w_i \quad {\rm minimizes} \quad \|\bm x_i - \Phi\bm w_i\|,
\end{align*}
where $\|\cdot\|$ denotes the Euclidean norm. The solution to the above optimization problem is explicitly given by 
\begin{align}\label{eq:mnlse}
    \widehat{\bm w}_i = (\Phi^\top\Phi)^{\dagger} \Phi^\top \bm x_i,
\end{align}
where $(\Phi^{\top}\Phi)^{\dagger}$ denotes the Moore-Penrose pseudo-inverse matrix~\citep[e.g.,][]{Banerjee2014} of $\Phi^{\top}\Phi$. Using the estimated coefficients $\widehat{\bm w}_i$, we express the functional representation of the $i$-th subject's data as $x_i(t) = \widehat{\bm w}_i^\top\bm\phi(t)$. 

Regarding the choice of the number of basis functions $K$, traditional approaches often select $K$ to be smaller than the number of observations $M_i$ to avoid overfitting \citep{ramsay2005functional}. However, recent theoretical evaluations by \cite{hastie2022surprises} suggest that using a larger number of parameters (bases, in this context) can be beneficial in cases where the noise level is low and the model is misspecified. In light of these insights, we propose using an excess number of basis functions, combined with the minimum norm interpolator, for functionalization in FDA. This approach has the potential to capture more complex patterns in the data and improve the accuracy of interpolations or subsequent analyses, especially in low-noise settings or when the true underlying function does not perfectly align with the chosen basis.

\section{Functional Regression Model} \label{sec:method}
In this section, we construct estimators through basis expansions for three standard models.
\subsection{Scalar on Function Regression} \label{sec:SonF}
Consider an independently and identically distributed dataset $\mD:=\{x_i,y_i\}_{i=1}^N$, with explanatory function $x_i(\cdot)\in L_2(\mS)$ on domain $\mS\subset \R$ and scalar response variable $y_i\in\R$. Suppose that predicting the response $y$ when a new $x$ is observed is of interest. We employ the following scalar-on-function regression model~\citep[SonF,][]{hastie1993statistical,muller2005functional,araki2009functional}: 
\begin{align}\label{eq:SonF}
     y_i =\int_{\mS} x_i(s)\beta(s)ds +\varepsilon_i,
\end{align}
where $\beta\in L_2(\mS)$ is a functional coefficient and $\varepsilon_i$ is an error term with mean zero and finite variance. This model assumes a linear relationship between the functional predictor $x_i$ and the scalar response $y_i$, mediated by the functional coefficient $\beta$.

We can represent $x_i(s)$ and $\beta(s)$ using basis expansions:
\begin{align*}
    x_i(s) &= \sum_{k=1}^K w_{ik} \phi_k(s)  ,~~\mathrm{and}~~ \beta(s)= \sum_{k=1}^K b_k \phi_k(s), 
\end{align*}
where $\phi_k$ are the basis functions, $w_{ik}$ and $b_k$ are corresponding coefficients for $x_i$ and $\beta$, respectively, and $K$ is the number of basis functions. The coefficients $w_{ik}$ are obtained using the minimum norm interpolator~\eqref{eq:mnlse}; therefore, the $w_{ik}$ are known here. For notational simplicity, we write the above expansion in vector form as 
\begin{align}\label{eq:base}
     x_i(s) = \bm{w}_{i}^{(K)\top} \bm{\phi}^{(K)}(s) ,~~\mathrm{and}~~ \beta(s)= \bm{b}^{(K)\top} \bm{\phi}^{(K)}(s) ,
\end{align}
where $\bm{\phi}^{(K)}(s) :=(\phi_1(s),\ldots,\phi_K(s))^{\top}$, $\bm{w}_{i}^{(K)} := (w_{i1},\ldots,w_{iK})^{\top}$ and $\bm{b}^{(K)} := (b_{1},\ldots,b_{K})^{\top}$. The upper subscripts of the vectors are added to explicitly represent the number of bases.

Using the above expansion, we can rewrite \eqref{eq:SonF} as
\begin{align}
     y_i &=  \bm{w}_{i}^{(K)\top} \Phi^{(K)} \bm{b}^{(K)} +\varepsilon_i \notag \\
         &=  \bm{z}_{i}^{\top}  \bm{b}^{(K)} +\varepsilon_i, \label{eq:approx:SonF}
\end{align}
where $\Phi^{(K)}$ denotes the $K\times K$ matrix, whose $(i,j)$-th entry is $\int_{\mS} \phi_i(s)\phi_j(s) ds$, and $\bm{z}_{i}= \Phi^{(K)}\bm{w}_{i}^{(K)}$. Then, the joint equation for all observations can be written as
\begin{align}
    \bm{y} = Z \bm{b}^{(K)} +\bm{\varepsilon},
    \label{eq:approx:SonF2}
\end{align}
where $\bm{y}=(y_1,y_2,\ldots,y_N)^{\top}$, $Z^{\top}=(\bm{z}_1^{\top},\bm{z}_2^{\top},\ldots,\bm{z}_N^{\top})^{\top}$ and $\bm{\varepsilon}=(\varepsilon_1,\varepsilon_2,\ldots,\varepsilon_N)^\top.$

When $K<N$, the ordinary least squares estimator $(Z^{\top}Z)^{-1}Z^{\top}\bm{y}$ can be used to estimate $\bm{b}^{(K)}$. However, we are interested in the case where $K$ can be larger than $N$, and $Z^{\top}Z$ is not invertible. Then, we introduce the minimum norm interpolator:
\begin{align*}
    \argmin_{\bm b^{(K)}}  \|\bm b^{(K)}\| \quad  {\rm s.t. } \quad \bm b^{(K)} \quad {\rm minimizes} \quad \|\bm y - Z\bm b^{(K)}\|,
\end{align*}
which is equivalent to
\begin{align} \label{eq:MNIP1}
    \hat{\bm{b}}^{(K)} = ( Z^{\top}Z)^{\dagger} Z^{\top} \bm{y}.
\end{align}
In other words, we adopt $Z\hat{\bm{b}}^{(K)}$ as the predictor of the new observations.

Since, in real measurements, data are observed at a finite number of discrete time points, we need to take that number into account. Here, for brevity, the number of observation points is assumed to be common across all individuals. Let $M$ be the number of $x$ observation points (it should be noted that the following discussion can be extended in a straightforward way to the case in which the number of observations is heterogeneous). Since $M$ controls the information contained in the regression model, it will have a significant impact on prediction accuracy.

Now, for precise prediction, we explore the way to select the number of bases, which is the only value that the analysts can control. To investigate the relationship between the number of basis functions $K$, the sample size $N$, and the number of observation points $M$, and their impact on the double-descent phenomenon, we consider two scenarios:
\begin{itemize}
    \item[(A)] $N<M$: If $1\leq K< M$, the model in~\eqref{eq:approx:SonF2} is a regression problem with sample size $N$ and number of parameters $K$. As $K$ gradually increases from 1, a double-descent phenomenon with a peak at $K=N$ will be observed. This can be understood by regarding the original regression as an over-parameterized linear regression.
    \item[(B)] $M<N$: In this case, since $\mathrm{rank} Z~(\le M)$ is less than $N$, the double-descent with respect to $N$ does not occur. Since the expressive power of the model in \eqref{eq:approx:SonF2} is limited to less than the number of observation points if $M$ is small, accuracy will reach a ceiling even when the number of bases is increased.
\end{itemize}

The model considered here is a simple linear regression model, and the concern in such a case is model misspecification. In real data analysis, the true functional data (i.e., the data generating process) is unknown, and there are features that cannot be captured by a finite set of basis functions chosen arbitrarily by the analyst. For example, approximating a function with a few dozen spline bases may not describe periodicity or the variation of spikes. In a rough sense, Equation~\eqref{eq:approx:SonF} is considered a misspecified model. However, as stated in Section~5 of \cite{hastie2022surprises}, even if the model is misspecified, increasing the dimension of the parameters will contribute to improved prediction accuracy. This implies that increasing the number of basis functions is also robust to model misspecification, providing further motivation for the use of excess basis functions in functional regression.

\subsection{Function on Function Regression} \label{sec:FonF}
Consider an independent and identically distributed dataset $\mD:=\{x_i, y_i\}_{i=1}^N$, where $x_i(\cdot)\in L_2(\mS)$ is an explanatory function on domain $\mS\subset \R$, and $y_i(\cdot)$ is a response function on domain $\mT\subset\R$. Our goal is to predict the response function $y$ when a new function $x$ is observed. We adopt the following function-on-function regression model~\citep[FonF,][]{ramsay1991some,matsui2009regularized}: 
\begin{align}\label{eq:FonF}
     y_i(t) =\int_{\mS} \beta(s,t)x_i(s)ds +\varepsilon_i(t),
\end{align}
where $\beta(s,t)$ is a bivariate functional coefficient, and $\varepsilon_i(t)$ is an error process with mean zero and constant variance $\sigma^2$. This model assumes a linear relationship between the functional predictor $x_i$ and the functional response $y_i$, mediated by the bivariate functional coefficient $\beta$.

Using basis expansion, as in Equation~\eqref{eq:base}, we can represent the functional predictor, the bivariate functional coefficient, and the functional response as
\begin{equation*}
    x_i(s) = \bm{w}_{i}^{(K_1)\top} \bm{\phi}^{(K_1)}(s) ,\ \ \ \beta(s,t)=\bm{\phi}^{(K_1)\top}(s) B \bm{\psi}^{(K_2)}(t),\ \ \  y_i(t) = \bm{v}_{i}^{(K_2)\top} \bm{\psi}^{(K_2)}(t) ,
\end{equation*}
where $\bm{v}_{i}^{(K_2)} = (v_{i1}, \ldots, v_{iK_2})^\top$ is the coefficient vector of the bases $\bm{\psi}^{(K_2)}(t) = (\psi_1(t), \ldots, \psi_{K_2}(t))^\top$, and $B$ is the coefficient matrix of $\bm{\phi}^{(K_1)}(s)$ and $\bm{\psi}^{(K_2)}(t)$. Here the coefficients $w_{ik}$ $(k=1, 2, \ldots, K_1)$ and $v_{il}$ $(l=1, 2,\ldots, K_2)$ are obtained using the minimum norm interpolator, as described in Equation~\eqref{eq:mnlse}. Substituting the basis function expansions into Equation~\eqref{eq:FonF}, we obtain
\begin{align}
     \bm{v}_{i}^{(K_2)\top} \bm{\psi}^{(K_2)}(t)  =  \bm{w}_{i}^{(K_1)\top} \Phi^{(K_1)} B \bm{\psi}^{(K_2)}(t)+\varepsilon_i(t).
\end{align}
To estimate the coefficient matrix $B$, we consider solving the following minimization problem:
\begin{align*}
    \argmin_{B\in \R^{K_1\times K_2}}  \|\vect (B)\| \quad  {\rm s.t. } \quad B \quad {\rm minimizes} \quad \|V \bm{\psi}^{(K_2)}(t)  -  Z B \bm{\psi}^{(K_2)}(t)\|_{L_2}, 
\end{align*}
where $V=(\bm{v}_{1}^{(K_2)},\bm{v}_{2}^{(K_2)},\ldots,\bm{v}_{N}^{(K_2)})^{\top}$, $\vect(\cdot)$ is the vectorization operator of a matrix and $\|\cdot \|_{L_2}$ is $L_2$ norm. Then, minimizing the least square error yields 
\begin{align}\label{est:FonF}
    \vect(\hat{B}) = (\Psi \otimes Z^{\top}Z)^{\dagger} \vect(Z^{\top}V\Psi),
\end{align}
where $\Psi$ is a $K_2\times K_2$ matrix whose $(i,j)$-th entry is $\int_{\mT} \psi_i(t)\psi_j(t) dt$. We consider this to be an estimator for the FonF problem.

In practice, the functional predictor and response are observed at a finite number of discrete time points. Let $M_1$ and $M_2$ be the number of time points for $x$ and $y$, respectively, assumed, for simplicity, to be the same across individuals. The dimensions of the observed data can affect the properties of the estimator. There are many possible combinations of the sample size $N$, the number of observation points $M_1$ and $M_2$, and the number of basis functions $K_1$ and $K_2$. However, two scenarios are particularly relevant to the double-descent phenomenon:
\begin{itemize}
    \item[(C)] $M_2$ and $K_2$: The parameter $K_2$ directly influences the prediction of the function $y$. Based on the idea that a function can be predicted with good accuracy if the unobserved parts are properly interpolated, increasing $K_2$ beyond $M_2$ may lead to the double-descent phenomenon in terms of prediction accuracy. In other words, the phenomenon can be attributed to the accuracy of the functionalization of the response.
    
    \item[(D)] $N$ and $K_1$: Following the same principle as (A) in the previous section, by increasing the number of basis functions for $x$ beyond the sample size $N$, a double-descent phenomenon can be observed as long as $M_1>N$. This corresponds to interpolating unobserved parts of the functional predictor using excess basis functions.
\end{itemize}

The double-descent phenomenon in FonF model can manifest in two ways: through the functionalization of the response (scenario C) and through the interpolation of the functional predictor (scenario D). By using excess basis functions in both the predictor and response expansions, we may be able to capture more complex patterns in the functional data and improve the accuracy of the functional regression model, even when the number of basis functions exceeds the number of observation points or the sample size. This further motivates the use of excess basis functions in functional regression settings.

\section{Numerical Experiments} \label{sec:num}

\subsection{SonF Regression}
As discussed at the end of Section~\ref{sec:SonF}, the accuracy of our predictions in SonF regression can be influenced by the various interrelationships among the sample size $N$, the number of observation points $M$, and the number of basis functions $K$. We investigated the prediction performances for scenarios (A) and (B) as described in Section~\ref{sec:SonF}. Table~\ref{tab:sumAB} summarizes the simulation settings. Although multiple criteria have been devised for basis selection, we conduct experiments with the number of bases selected through five-fold cross-validation~\citep[CV,][]{stone1974cross}, selected by corrected AIC~\citep[cAIC,][]{sugiura1978further,bedrick1994model}, and fixed at a value of $50$. Note that when cAIC is used, the error terms of the regression model are assumed to be independent Gaussian.
% パラメーター数はK

\begin{table}[t]
\centering
\caption{Summary of simulation settings and representations for scalar-on-function regression.}
\begin{tabular}{@{}>{\centering\arraybackslash}m{1.5cm}m{7cm}m{3cm}m{3cm}@{}}
\toprule
Symbol & Description & Scenario (A) & Scenario (B) \\
\midrule
$N$ & Size of training dataset & Variable & Fixed ($50$) \\
$N_{\text{test}}$ & Size of test dataset & Fixed ($150$) & Fixed ($150$) \\
$M$ & Number of measurements for $x$ & Fixed ($75$) & Variable \\
$K$ & Number of bases for $x$ & Variable & Variable \\
\bottomrule
\end{tabular}
\label{tab:sumAB}
\end{table}

\subsubsection*{Scenario (A)}
Consider the situation where the number of observation points $M$ is larger than the sample size $N$, discussed in Section~\ref{sec:SonF}. First, we present the data-generating process. The functions $x_i(s)$ and $\beta(s)$ are produced by Gaussian processes (GPs) with the radial basis function kernel \citep[RBF,][]{rasmussen2006gaussian} $k(x_1,x_2) = \theta^2 \exp(-\|x_1-x_2\|^2/h^2)$, whose hyperparameters are set to $(\theta,h)=(10,10)$ and $(15,10)$, respectively. The generated $x_i(s)$ are then centered to have a mean of $0$. We then generate $y_i$ by adding a standard normal noise to the integral of the product of $x_i(s)$ and $\beta(s)$. The observation vectors $\{\bm{x}_i\}$ are derived by selecting $M=75$ random points from the functions plus a standard normal noise $N(\bm{0}, I_M)$. We set the training data size to $N=5,10$ and $20$.

For each $N$, we used the above procedure to generate $50$ datasets, each with $N$ observations as a training set and $150$ data points as a test set, and then analyzed each dataset using natural splines~\citep{wood2017generalized, Rteam} and \eqref{eq:MNIP1}. Specifically, for $N$ observations, we calculated \eqref{eq:MNIP1}, varying the number of bases $K$ from $4$ to $50$. To assess the performance of the model, we computed the mean squared error (MSE) of the predictions from the true signal for the $150$ test data and analyzed the changes in MSE as $K$ increased.

\begin{table}[t]
\centering
\caption{MSEs of scalar-on-function regressions for different basis selection methods, averaged over $50$ simulated datasets.\label{tab1}}
\begin{tabular}{l ccc ccc}
\toprule
 & \multicolumn{3}{c}{Scenario (A)} & \multicolumn{3}{c}{Scenario (B)} \\
\cmidrule(lr){2-4} \cmidrule(lr){5-7}
Method     & {$N=5$} & {$N=10$} & {$N=20$}  & {$M=5$} & {$M=10$} & {$M=20$} \\
\midrule
CV     & 21.992  &  9.650  & 8.654   & 27.549 & 18.852 & 4.833  \\
Fixed  & 39.805  & 20.473  & 8.853   & 27.549 & 18.874 & 5.120  \\
cAIC   & 54.452  & 22.387  & 8.950   & 28.608 & 18.882 & 5.490  \\
\bottomrule
\end{tabular}
\end{table}

\begin{figure}[t]
    \centering
    \includegraphics[width=0.45\textwidth]{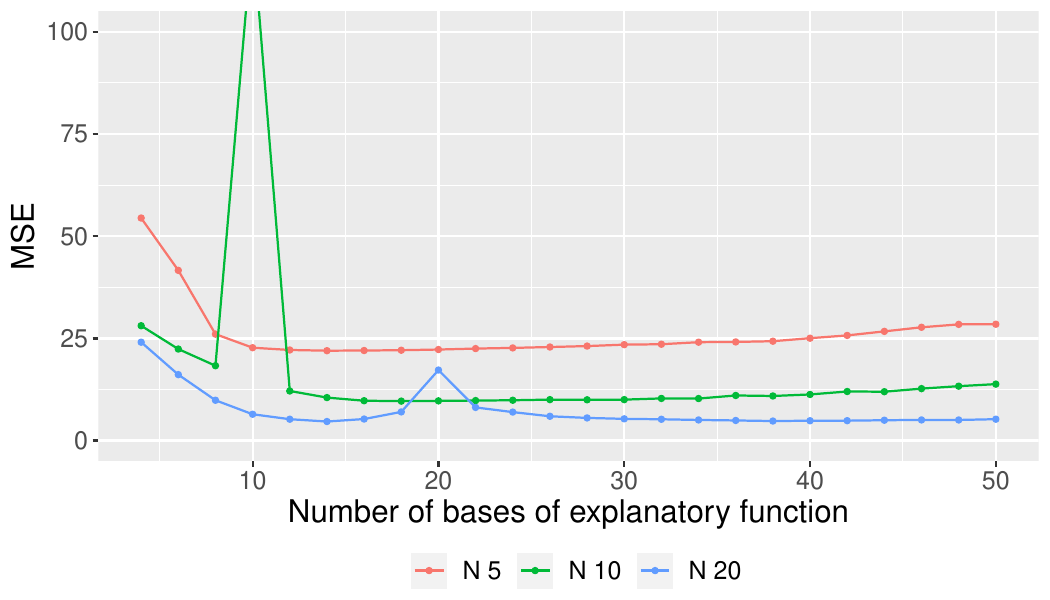}
    \includegraphics[width=0.45\textwidth]{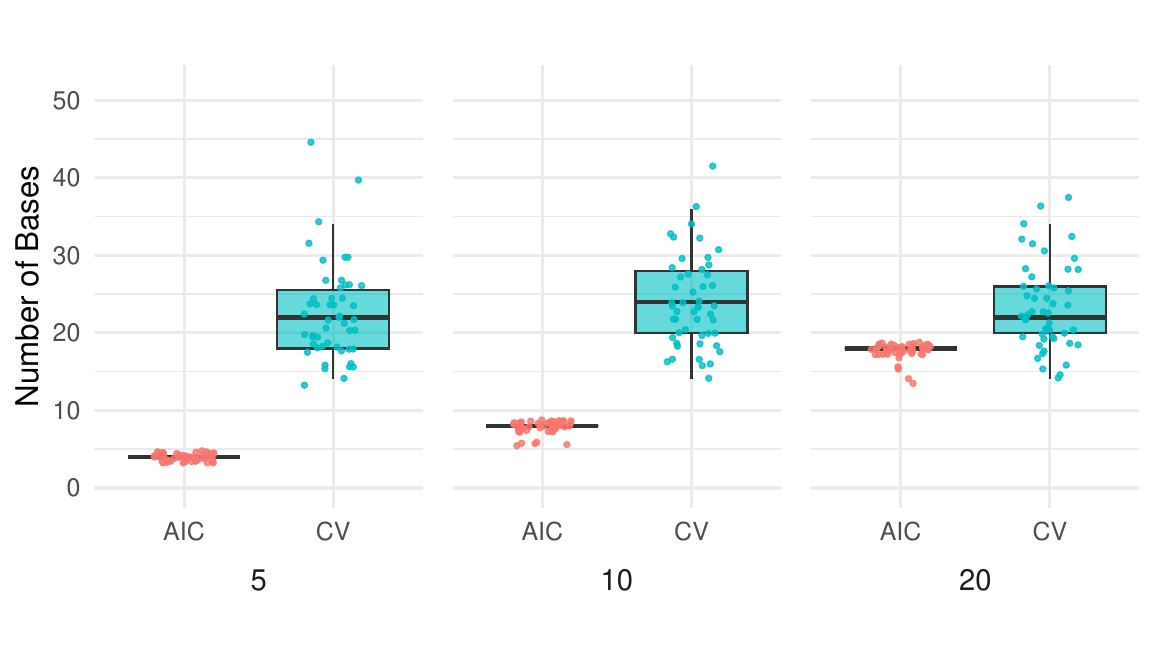}
    \caption{Left: MSE for varying number of bases ($K$) and sample size ($N$) in Scenario (A). Right: Box plots showing the number of bases selected by AIC and 5-fold cross-validation (CV) in Scenario (A).  \label{fig:SonF1}}
\end{figure}

The left panel in Figure~\ref{fig:SonF1} illustrates, for one representative dataset, how the number of bases $K$ affects predictions when $M$ is large. Initially, the MSE increases rapidly as $K$ approaches the sample size $N$; however, it peaks and begins to decrease when $K$ becomes larger than $N$, exhibiting the double-descent phenomenon. Next, observe the quantitative evaluation in Table~\ref{tab1}, whose entries represent the average values of the MSEs over $50$ datasets. Note that since cAIC assumes a situation where the degrees of freedom are smaller than $N$, the optimal number of basis functions selected is found before the peak. However, the prediction accuracy of the predictor with a fixed number of basis functions ($K=50$) is superior to the case in which the number of basis functions is selected using cAIC. For CV, which solely considers the goodness of fit of the predictions, the prediction accuracy after the peak is better than before the peak, as can be seen in the right panel of Figure~\ref{fig:SonF1}. These findings indicate that choosing a number of basis functions that is larger than the sample size is preferable in this scenario.

\begin{figure}[t]
    \centering
    \includegraphics[width=0.45\textwidth]{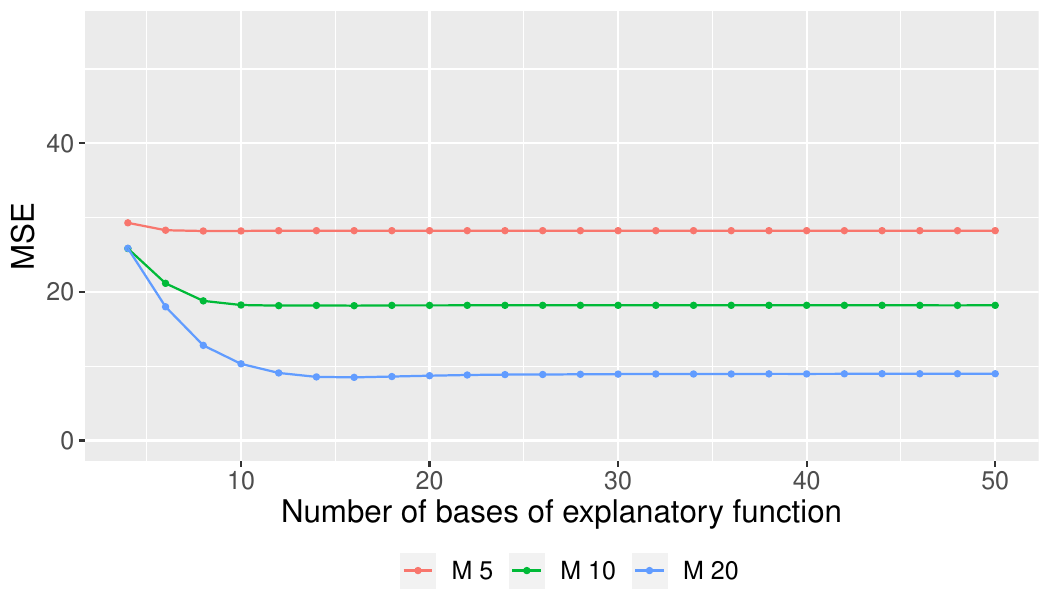}
    \includegraphics[width=0.45\textwidth]{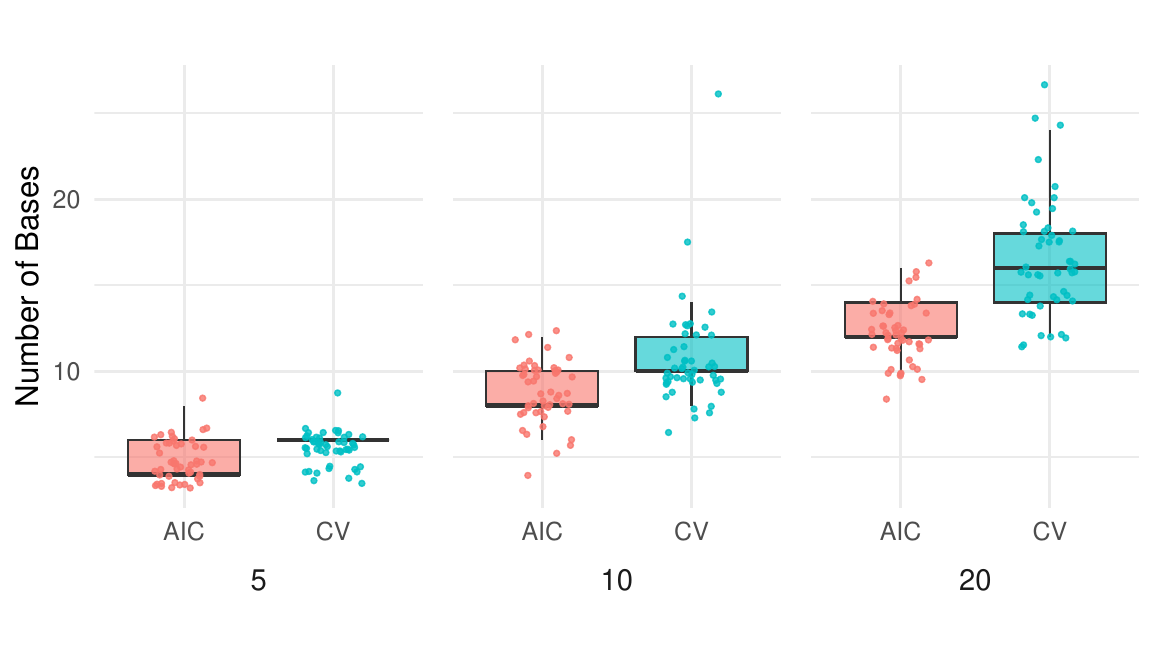}
    \caption{Left: MSE for varying numbers of bases ($K$) and observation points ($M$) in Scenario (B). Right: Box plots showing the number of bases selected by AIC and 5-fold cross-validation (CV) in Scenario (B).  \label{fig:SonF2}}
\end{figure}

\subsubsection*{Scenario (B)}
Next, we focus on the situation where the number of observation points $M$ is smaller than the sample size $N$. The functions $\bm{x}_i$ and $\beta$ and the response $y_i$ were generated in the same manner as in the previous scenario. In this setting, we generated $N=50$ data points for the training dataset and $150$ data points for the test dataset, with $M$ taking on the values $5$, $10$, and $20$.

We produced $50$ datasets through the above procedure and analyzed each. For each value of $M$, in each of the datasets, we trained the parameters using \eqref{eq:MNIP1}, with $K$ natural spline bases ($K$ varied from $4$ to $50$), on the training data and then calculated the MSE on the test data to examine how the MSE values changed as the number of basis functions $K$ increases.

The results are displayed in Figure~\ref{fig:SonF2} and Table~\ref{tab1}, where the reported values are averaged over the $50$ datasets. In this scenario, the rank of the design matrix in \eqref{eq:approx:SonF2} is low, which implies that the degrees of freedom of the model remain unchanged even as the number of basis functions increases. The right panel in Figure~\ref{fig:SonF2} shows that CV did not choose an excess number of bases. As a result, the observed MSEs ceased to decrease at around $K=M$, suggesting that increasing the number of basis functions beyond this point is not particularly advantageous. Hence, if the number of observation points restricts the expressive power of the model of the regression model, the double-descent phenomenon does not occur.

These simulation studies demonstrate the potential benefits of using excess basis functions in SonF regression when the number of observation points is sufficiently large (Scenario A). The double-descent phenomenon is clearly observed, with the prediction accuracy improving as the number of basis functions increases beyond the sample size. However, when the number of observation points is limited (Scenario B), increasing the number of basis functions beyond the number of observation points does not lead to further improvements in prediction accuracy, and the double-descent phenomenon is not observed. These findings highlight the importance of considering the interplay between the sample size, the number of observation points, and the number of basis functions when applying scalar-on-function regression in practice. The use of excess basis functions, combined with the minimum norm interpolator, can be a valuable approach for improving prediction accuracy in scenarios where the number of observation points is sufficiently large relative to the sample size.

\subsection{FonF Regression}
As discussed in Section~\ref{sec:FonF}, the prediction accuracy in FonF regression is influenced by the interplay between the sample size $N$, the number of observation points for the predictor and response functions ($M_1$ and $M_2$), and the number of basis functions for the predictor and response functions ($K_1$ and $K_2$). We now demonstrate Scenarios (C) and (D) through the following numerical experiments. The settings are summarized in Table~\ref{tab:sumCD}.
% パラメーター数はK_1 \times K_2

\begin{table}[t]
\centering
\caption{Summary of simulation settings and symbols for function-on-function regression.}
\label{tab:sumCD}
\begin{tabular}{@{}>{\centering\arraybackslash}m{1.5cm}m{7cm}m{3cm}m{3cm}@{}}
\toprule
Symbol & Description & Scenario (C) & Scenario (D) \\
\midrule
$N$ & Size of training dataset & Fixed ($50$) & Variable \\
$N_{\text{test}}$ & Size of test dataset & Fixed ($150$) & Fixed ($150$) \\
$M_1$ & Number of measurements for $x$ & Fixed ($75$) & Fixed ($75$) \\
$M_2$ & Number of measurements for $y$ & Variable & Fixed ($75$) \\
$K_1$ & Number of bases for $x$ & Fixed ($10$) & Variable \\
$K_2$ & Number of bases for $y$ & Variable & Fixed ($10$) \\
\bottomrule
\end{tabular}
\end{table}

\subsubsection*{Scenario (C)}
Here, we investigate the relationship between $K_2$ (number of basis functions for the response function $y$) and $M_2$ (number of observation points for $y$). We consider the scenario where both the predictor $x$ and the response $y$ are functions. Specifically, we sampled $x$ from a GP whose kernel is an RBF  having hyperparameters $(\theta, h) = (10, 10)$ and centered it to be zero-mean. For every $t$, we sampled $\beta(\cdot,t)$ from a GP with an RBF kernel having hyperparameters $(\theta, h) = (15, 10)$.  The true response function was generated by integrating the product of $\beta(s,t)$ and $x_i(t)$ as \eqref{eq:FonF}, and the observations $\{\bm{y}_i\}$ were given by adding standard normal noise to $M_2$ points extracted from the function. Moreover, the observation vectors $\{\bm{x}_i\}$ are derived by randomly selecting $M_1=75$ points from the functions and adding standard normal noise. For each $M_2=5,10$ and $20$, we generated $N=50$ observations as a training set and $150$ values as a test set.

For each value of $M$, we generated $50$ datasets using the above procedure and analyzed each dataset using natural splines and \eqref{est:FonF} on the training sample of size $N$, fixing $K_1$ at $10$ and varying $K_2$ from $4$ to $50$. We then examined the relationship between the number of basis functions $K_2$ of the response function and the MSE for the test data.

\begin{table}[t]
\centering
\caption{MSEs of function-on-function regressions for different basis selection methods, averaged over $50$ simulated datasets.\label{tab2}}
\begin{tabular}{l ccc ccc}
\toprule
 & \multicolumn{3}{c}{Scenario (C)} & \multicolumn{3}{c}{Scenario (D)} \\
\cmidrule(lr){2-4} \cmidrule(lr){5-7}
Method     & {$M_2$=5} & {$M_2$=10} & {$M_2$=20}  & {$N=5$} & {$N=10$} & {$N=20$} \\
\midrule
CV     & 9.263 & 8.652 & 10.744  & 8.021  & 5.257  & 3.500   \\
Fixed  & 9.976 & 9.018 & 10.881  & 8.033  & 5.399  & 3.502   \\
cAIC   & 337.620 & 99.037 & 11.761  & 18.746  & 9.602  & 8.214  \\
\bottomrule
\end{tabular}
\end{table}

\begin{figure}[t]
    \centering
    \includegraphics[width=0.45\textwidth]{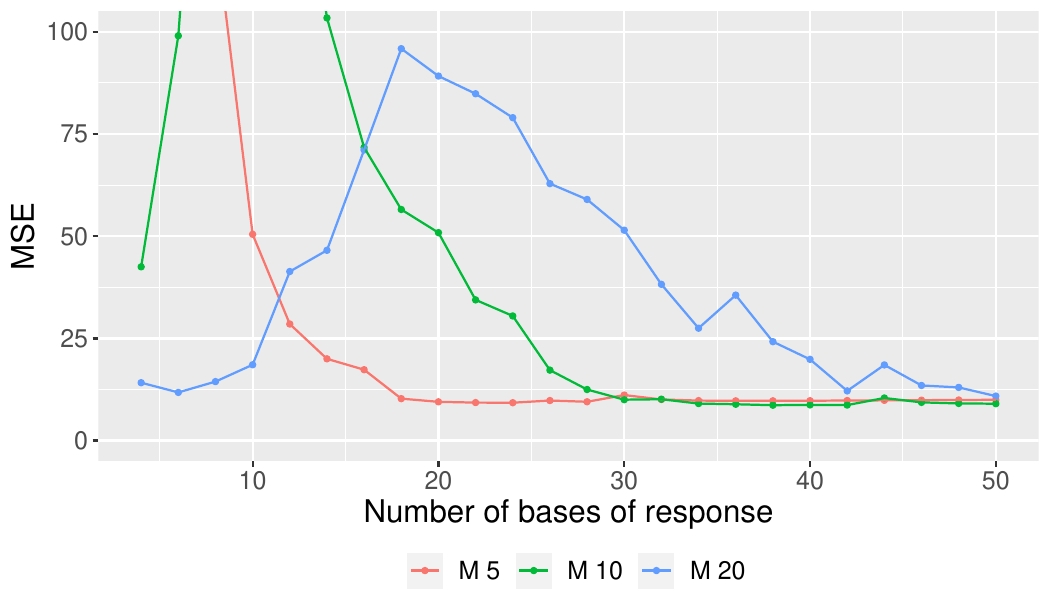}
    \includegraphics[width=0.45\textwidth]{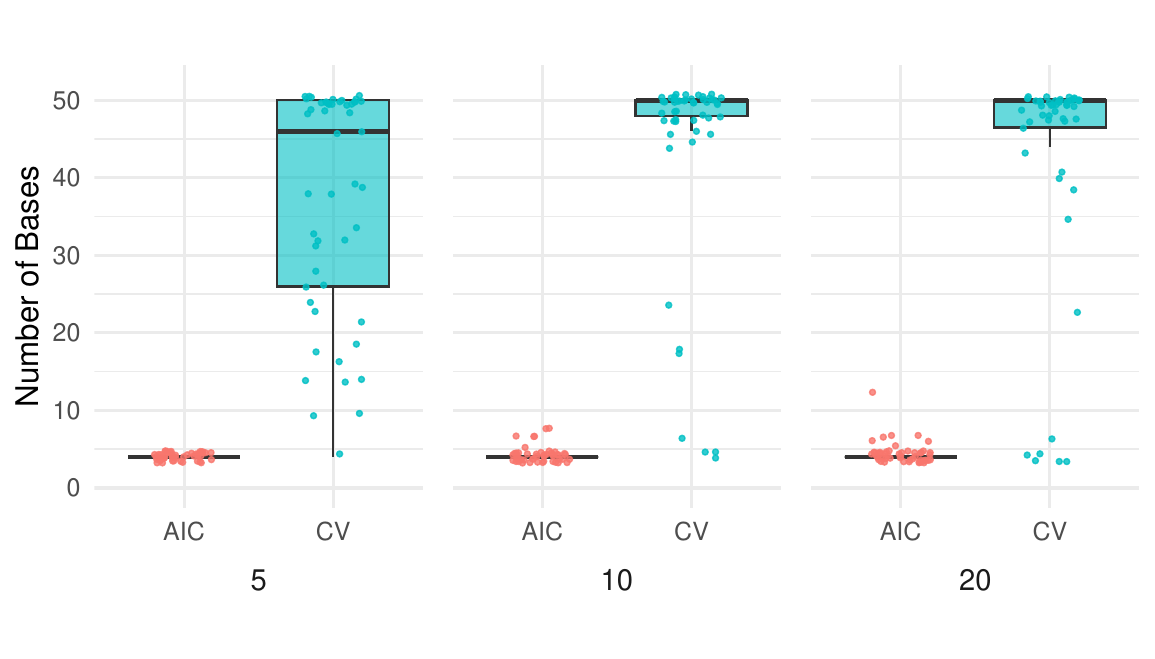}
    \caption{Left: MSE for varying numbers of bases for the response ($K_2$) and varying numbers of measurements for the response ($M_2$) in scenario (C). Right: Box plots showing the number of bases selected by AIC and 5-fold cross-validation (CV) in Scenario (C).  \label{fig:FonF1}}
\end{figure}

\begin{figure}[t]
    \centering
    \includegraphics[width=0.45\textwidth]{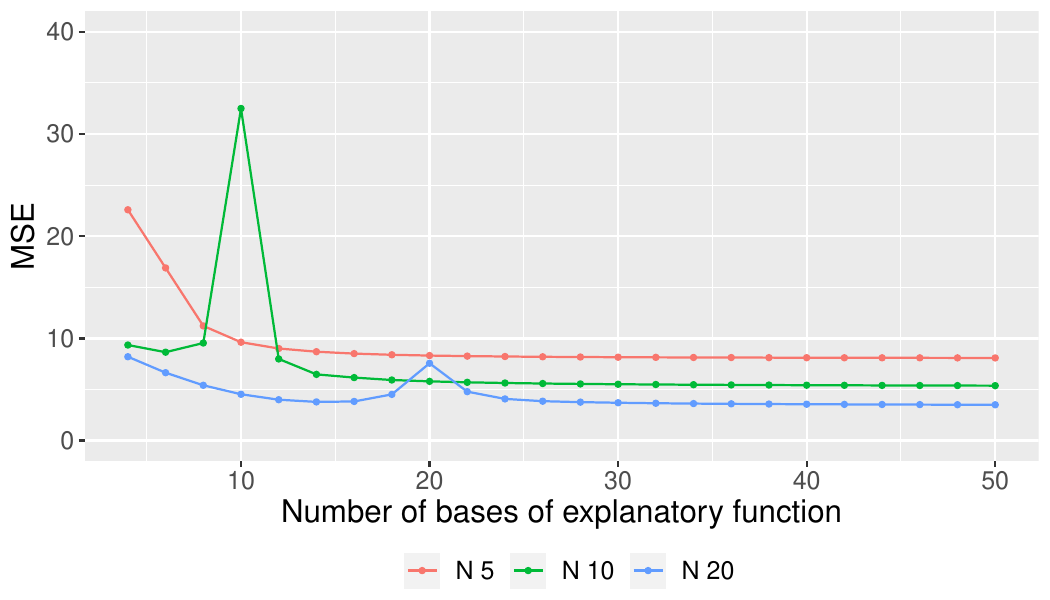}
    \includegraphics[width=0.45\textwidth]{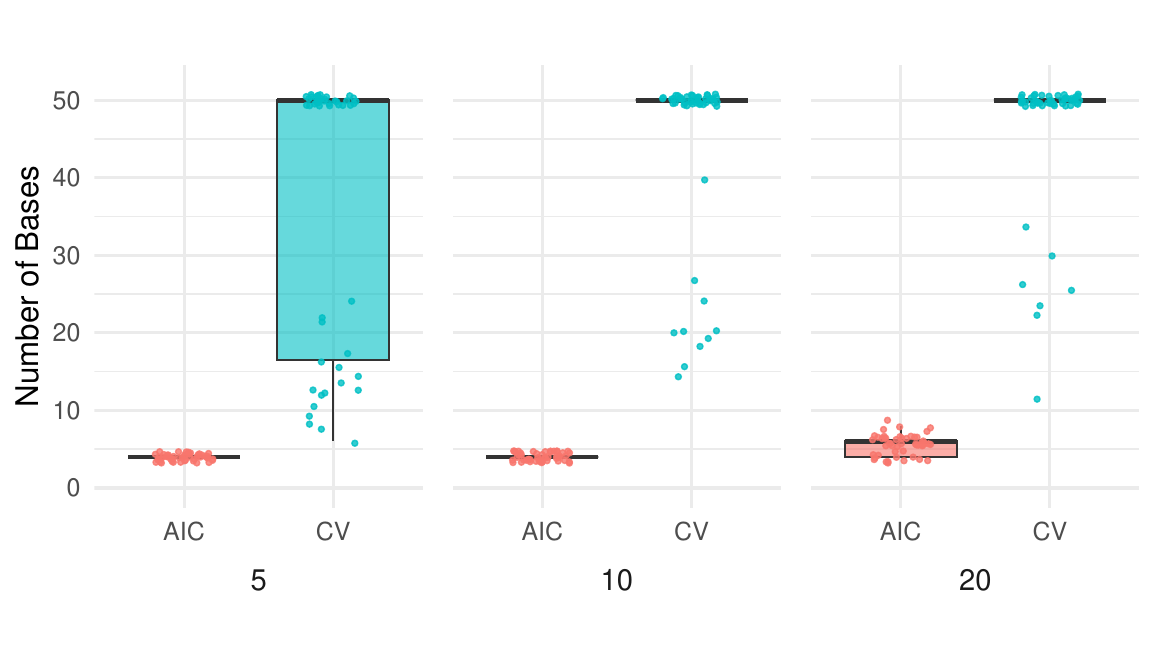}
    \caption{Left: MSE for varying numbers of bases for the predictor ($K_1$) and training sample sizes ($N$) in scenario (D). Right: Box plots showing the number of bases selected by AIC and 5-fold cross-validation (CV) in Scenario (D).  \label{fig:FonF2}}
\end{figure}

The left panel in Figure~\ref{fig:FonF1} illustrates the change in the MSE values with the increasing number of bases for a representative dataset. The results show that the MSE value reaches its maximum when the number of bases of the response function equals the size of the training sample, after which the MSE decreases. As the individual prediction targets are functions, the prediction (interpolation of the predicted function) improves with an increase in the number of bases. Table~\ref{tab2} shows the results when $K_2$ is selected by CV and cAIC, respectively. As indicated, the basis selection via cAIC results in poor prediction performance. In particular, $M_2=5$ and $M_2=10$ fail to predict the response function either because the number of bases is too small to represent the function or because of overfitting. This poor performance can be attributed to the fact that the basis of the response function itself is considered, suggesting that the choice of basis is particularly sensitive in this scenario. In contrast, CV, choosing a large number of bases (the right panel in Figure~\ref{fig:FonF1}) or fixing the number of bases at large values contributes to good interpolation performance.

\subsubsection*{Scenario (D)}
In this section, we examine the relationship between $K_1$ (number of basis functions for the predictor function $x$) and $N$ (sample size). The generating process for the functions $x$, $\beta$, and $y$ in Equation~\eqref{eq:FonF} is the same as in the previous section. The observation vectors ${\bm{x}_i}$ and ${\bm{y}_i}$ are both obtained by randomly selecting $75$ points from the functions $x_i$ and $y_i$, respectively, and adding centered Gaussian noise with unit variance. For $N=5,10$, and $20$, we generated $N$ observations as the training set and $150$ observations as the test set.

For each $N$, we generated $50$ datasets and analyzed each one using \eqref{est:FonF}, varying $K_1$ from $4$ to $50$ and fixing $K_2=10$ natural spline basis functions. We investigated the relationship between the number of basis functions for the predictor function $K_1$ and the MSE for different training sample sizes $N$.

The simulation results are given in Figure~\ref{fig:FonF2} and Table~\ref{tab2}. The left panel in Figure~\ref{fig:FonF2} illustrates the change in MSE with sample size $N$ for a representative dataset. Once again, the double-descent phenomenon is observed in this scenario. This result is essentially the same as in Scenario (A), as it involves the relationship between the sample size and the number of basis functions for the predictor function (although the number of observation points $M_1$ must be greater than $N$). The right panel in Figure~\ref{fig:FonF2} shows that CV tends to select excess bases; Table~\ref{tab2} confirms that, as before, using a larger number of basis functions results in better prediction accuracy.

These simulation results highlight the benefits of using excess basis functions in the FonF model for both the response function (Scenario C) and the predictor function (Scenario D). The double-descent phenomenon is evident in both scenarios, with the prediction accuracy improving as the number of basis functions increases beyond the sample size or the number of observation points for the response function. These findings underscore the practical importance of considering the interplay between the sample size, number of observation points, and number of bases.

\section{Application to real datasets}\label{sec:app}
This section provides examples of the double-descent phenomenon in functional regression, as evidenced by empirical data. We examine Scenario (A) across two commonly used datasets.

\subsection{Gasoline Dataset}
First, we focused on the ``gasoline'' dataset, stored in the \texttt{R} language ``refund'' package~\citep{refund}. This dataset comprises octane numbers for 60 gasoline samples and their near-infrared reflectance spectra. The octane number serves as a scalar indicator, quantifying the combustion quality of the gasoline, and the 401 near-infrared reflectance spectra represent the molecular structure of the substance. 

In this analysis, following \cite{reiss2007functional} and \cite{reiss2009smoothing}, we treated a set of near-infrared reflectance spectra as a functional explanatory variable and considered the problem of predicting the octane number, treated as a response, based on the minimum norm interpolator \eqref{eq:MNIP1}, varying the number of natural spline basis functions We randomly selected 10 observations as the training data and calculated the MSE of the predictions on the remaining 50 observations, which served as the test data.

\begin{figure}[t]
  \begin{center}
  \includegraphics[width=0.6\textwidth]{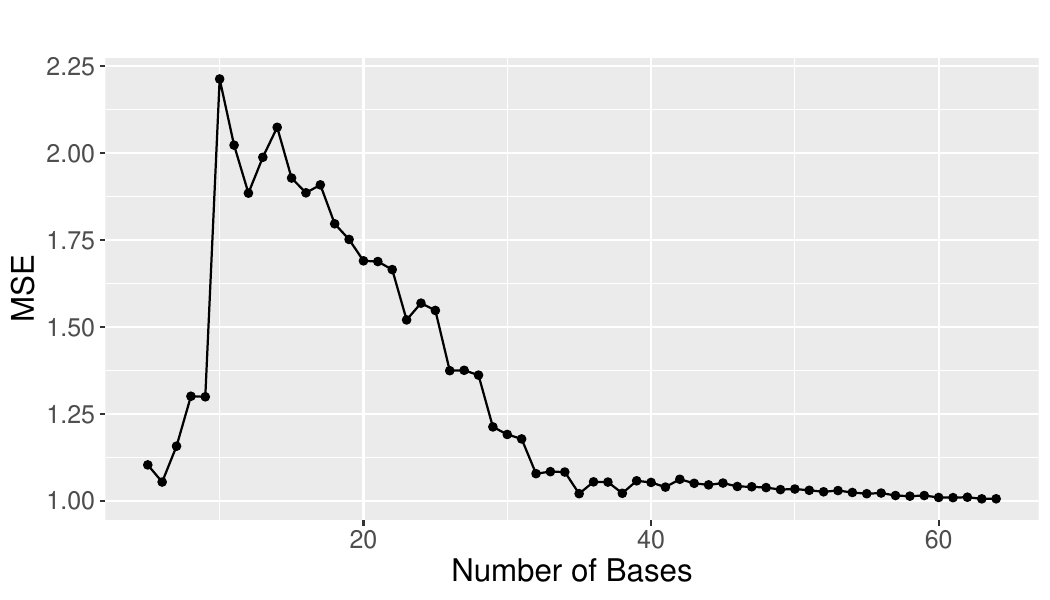}
  \end{center}
  \caption{Relationship between the number of bases and MSE for gasoline dataset.\label{fig:gasoline}}
\end{figure}

The MSEs with varying numbers of basis functions are shown in Figure~\ref{fig:gasoline}. As can be seen in the figure, the MSE peaks at the same point as the size of the training sample ($10$) and then gradually decreases. Notably, when the number of basis functions exceeds 50, the MSE becomes smaller than when fewer basis functions are used. This outcome suggests that leveraging a large number of basis functions can indeed enhance prediction accuracy for real data, as evidenced by the double-descent phenomenon shown here.

\subsection{Diffusion Tensor Imaging Dataset}
Next, we address the diffusion tensor imaging (DTI) dataset, which is commonly used in functional data analysis and is stored as ``DTI'' in the \texttt{R} language ``refund'' package. DTI is a modality based on magnetic resonance imaging (MRI) that allows the diffusion of water in the brain to be tracked. One hundred patients are scanned for DTI approximately once a year and undergo the PASAT (Paced Auditory Serial Addition Test), a neuropsychological test used to assess cognitive function.

Within this framework, following \cite{goldsmith2011penalized} and \cite{goldsmith2012longitudinal}, we considered the fractional anisotropy tract profiles of the corpus callosum area (CCA) as a functional explanatory variable to predict the subject's PASAT score as a response. Although patients may visit the clinic multiple times, each visit is treated as a distinct data point; data with missing values were removed. This resulted in a sample size $N=334$, with $93$ observation points (i.e., $M=93$) for the explanatory variable CCA. We performed predictions based on \eqref{eq:MNIP1} with natural spline bases, varying the number of bases. A training sample of size 20 was used, with the remaining 314 observations serving as the test data.

\begin{figure}[t]
  \begin{center}
  \includegraphics[width=0.6\textwidth]{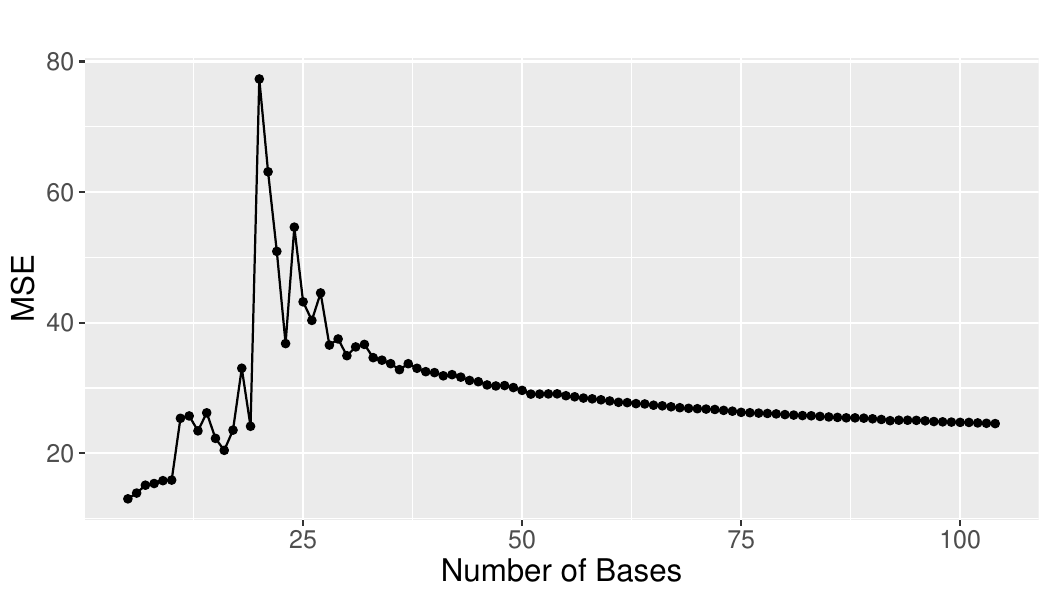}
  \end{center}
  \caption{Relationship between the number of bases and MSE for the DTI dataset. \label{fig:DTI}}
\end{figure}

As illustrated in Figure~\ref{fig:DTI}, the double-descent is evident. The MSE peaks at approximately the same value as the training sample size and drops smoothly from there. In this case, the MSE does not decrease as much as in the gasoline dataset, possibly because the functional form of the explanatory variable is simple, and a few basis functions are sufficient to represent the function. However, the double-descent phenomenon clearly occurs, indicating the risk of conventionally searching solely for a smaller number of basis functions than the size of the training sample based on the idea of preventing overfitting.

\section{Discussion} \label{sec:con}
This study questions the conventional notion that the number of basis functions should be smaller than the number of observation points and asserts the benefits of considering an excess number of basis functions in the FDA. In Section~\ref{sec:method}, we argue that in functional regression, if one uses a number of basis functions above a certain threshold, the double-descent phenomenon can be observed and better prediction accuracy can potentially be achieved. We demonstrated this phenomenon through numerical experiments and found that optimal prediction accuracy can be realized to the right of the peak of the double-descent curve. Importantly, this phenomenon is not merely the subject of theoretical analysis or numerical experiments but can also be observed in real-world datasets. In both the gasoline and DTI datasets, a clear double-descent was observed, with the gasoline dataset producing optimal prediction accuracy beyond the peak. These findings provide valuable guidance in the analysis of functional data, strongly suggesting that when selecting the number of basis functions, one should consider a wider range of possibilities and not be limited by the sample size or the number of observation points.
 
Future research should extend investigations of the practicality of this phenomenon to different types of datasets and models, including functional time series. Additionally, beyond the minimum norm interpolator, the advantage of excess basis functions may be further supported by ridge regression, although this would require tuning parameter selection. Moreover, the theoretical foundations of the double-descent phenomenon in functional data analysis should be more deeply explored. While this study provided empirical evidence and intuitive explanations, a rigorous mathematical analysis of the conditions under which the phenomenon occurs and its relationship to the properties of the functional data and the chosen basis functions would strengthen the understanding and applicability of our findings.

\section*{Computer Programs}
The computer programs used in this manuscript to demonstrate the double-descent curve in Section~\ref{sec:num} and the application presented in Section~\ref{sec:app} have been developed for execution in the \texttt{R} statistical computing environment. These programs are publicly available at the GitHub repository: \href{https://github.com/TomWaka/DD-FDR}{https://github.com/TomWaka/DD-FDR}.

\section*{Acknowledgements}
T. Wakayama was supported by JSPS KAKENHI (22J21090) and H. Matsui was supported by JSPS KAKENHI (23K11005). 

\bibliographystyle{plainnat}
\bibliography{ref}

\end{document}